\documentclass[11pt]{article}

\usepackage{amsmath}
\usepackage{amssymb}
\usepackage{amsfonts}
\usepackage{latexsym}
\usepackage{color}
\usepackage{bm}
\usepackage{cite}

\normalbaselines
\catcode `\@=11
\@addtoreset{equation}{section}

\catcode`\@=12

\newtheorem{theo}{Theorem}[section]

\newtheorem{lem}[theo]{Lemma}
\newtheorem{prop}[theo]{Proposition}
\newtheorem{cor}[theo]{Corollary}

\newtheorem{definition}[theo]{Definition}

\newtheorem{example}[theo]{Example}
 
\newtheorem{remark}[theo]{Remark}

\newcommand{\betheo}{\begin{theo}$\!\!\!${\bf } }
\newcommand{\entheo}{\end{theo}}

\newcommand{\becor}{\begin{cor}$\!\!\!$  }
\newcommand{\encor}{\end{cor}}

\newcommand{\belem}{\begin{lem}$\!\!\!$  }
\newcommand{\enlem}{\end{lem}}

\newcommand{\beprop}{\begin{prop}}
\newcommand{\enprop}{\end{prop}}

\newcommand{\bedefi}{\begin{definition}$\!\!\!$ \rm }
\newcommand{\findefi}{ \end{definition}}

\newcommand{\beex}{\begin{example}$\!\!\!$ \rm }
\newcommand{\enex}{ \end{example}}

\newcommand{\berem}{\begin{remark}$\!\!\!$ \rm }
\newcommand{\enrem}{ \end{remark}}

\newcommand{\qed}{\hfill \raisebox{-0.55mm}{{$\scriptstyle\square$}} \hspace*{-4.25mm}{$\scriptstyle\blacksquare$} \medskip \normalsize}

\newcommand{\be}{\begin{equation}}
\newcommand{\en}{\end{equation}}

\newcommand{\bea}{\begin{eqnarray}}
\newcommand{\ena}{\end{eqnarray}}

\newcommand{\beano}{\begin{eqnarray*}}
\newcommand{\enano}{\end{eqnarray*}}

\newcommand{\bee}{\begin{enumerate}}
\newcommand{\ene}{\end{enumerate}}

\newcommand{\bei}{\begin{itemize}}
\newcommand{\eni}{\end{itemize}}

\newcommand{\betab}{\begin{tabular}}
\newcommand{\entab}{\end{tabular}}

\newcommand{\up}{\raisebox{0.7mm}{$\upharpoonright $}}%

\newcommand{\ha }{^{\rm\textstyle *}}

\newcommand{\ov}[1]{\overline{#1}}

\newcommand{\mc}{\mathcal}
\newcommand{\mb}{\mathbb}
\newcommand{\RN}{\mb R}
\newcommand{\CN}{\mb C}
\def\NN{{\mathbb N}}
\def\ZN{{\mathbb Z}}

\def\D{{\mathcal D}}

\def\G{{\mathcal G}}
\def\H{{\mathcal H}}

\def\K{{\mathcal K}}

\def\P{{\mathcal P}}

\def\T{{\mathcal T}}
\newcommand{\norm}[2]{\left\| #2 \right\|_{#1}}
\newcommand{\dis}{\displaystyle}

\newcommand{\noi}{\noindent}
\newcommand{\ud}{\,\mathrm{d}}
\renewcommand{\leq}{\leqslant}
\renewcommand{\geq}{\geqslant}

\newcommand{\BH}{{\mc B}(\H)}

\def\hs{Hilbert space}

\newcommand{\ip}[2]{\langle {#1} |{#2} \rangle}

\def\OL{\relax\ifmmode {\sf L}\else{\textsf L}\fi}
\def\OR{\relax\ifmmode {\sf R}\else{\textsf R}\fi}

\newcommand{\bdim}{ {\bf Proof. }}
 \newcommand{\edim}{\qed}
\newcommand{\pip}{{\sc pip}-space}

\newcommand{\HG}{\H(G)}
\newcommand{\s}{\underline}

\begin{document}

%

\thispagestyle{empty}

\vspace*{1cm}
\begin{tabular}{l}
{\Large \textbf  {Some remarks on quasi-Hermitian operators }}

\vspace{1cm}
\\
 {\sf Jean-Pierre Antoine} \footnotemark 
\\ 
{\small\it Institut de Recherche en Math\'ematique et  Physique}\\ 
{\small\it Universit\'e Catholique de Louvain}\\
{\small\it  B-1348  Louvain-la-Neuve, Belgium} \bigskip
\\
{\sf Camillo Trapani} \footnotemark 
\\
{\small\it Dipartimento di Matematica e Informatica, }\\ 
{\small\it  Universit\`a di Palermo}\\
{\small\it  I-90123, Palermo, Italy}\bigskip
\end{tabular}

\vspace{1cm}
\begin{abstract}
A quasi-Hermitian operator  is an operator that is similar to its adjoint in some sense, via a
metric operator, i.e., a strictly positive self-adjoint operator. Whereas those metric operators are in general assumed to be bounded, we analyze the structure generated by unbounded metric operators in a Hilbert space.  Following our previous work, we introduce several generalizations of the notion of similarity between operators. Then  we explore systematically the various types of quasi-Hermitian operators, bounded or not. Finally we discuss their application    in the so-called  pseudo-Hermitian quantum mechanics.

\bigskip\bigskip\bigskip

 \footnotetext[1]{Electronic mail: {\tt jean-pierre.antoine@uclouvain.be}}
\footnotetext[2]{Electronic mail: {\tt camillo.trapani@unipa.it}}

\end{abstract}

\section{INTRODUCTION}
\label{sect_intro}

A long time ago, Dieudonn\'e \cite{dieudonne} introduced the term of quasi-Hermitian operators for characterizing those bounded operators $A$ which satisfy a relation of the form
\be\label{eq-quasiH1}
GA=A \ha  G,
\en
where $G$ is a \emph{metric operator}, i.e., a strictly positive self-adjoint operator. The same relation makes sense, however,  for unbounded operators $A$
also, under suitable conditions. In any case,
the operator $G$ then defines a new metric (hence the name) and a new Hilbert space (called physical in some applications), in which $A$ is symmetric and possesses a self-adjoint extension.
In particular, the Dieudonn\'e  relation implies that the operator $A$ is similar to its adjoint $A \ha $, in some sense, so that this notion of similarity plays a central  r\^ole in the theory.

In most of the literature, the metric operators are assumed to be bounded.  However, unbounded
metric operators have been  introduced in two recent works \cite{bag-zno2, mosta2}   in view of certain physical applications.
 In a recent paper \cite{pip-metric}  we have explored    the properties of unbounded metric operators, in particular, their incidence on similarity and on spectral data. We  will  continue this work here, mostly focusing on the various notions of similarity between operators. Whereas the \pip\ approach \cite{pip-book} was largely exploited in the first paper, here we will stick to standard operator theory in \hs s.

The notion of similarity and quasi-similarity between operators on Banach spaces has a long history, notably in the context of spectral operators, in the sense of Dunford \cite[Sec.XV.6]{dunford-schwartz}.  In particular, a  spectral operator of scalar type  is an operator that can be written as $A=\int_\CN \lambda \ud E(\lambda)$, where $E(\cdot)$ is a  bounded (but not necessarily self-adjoint) resolution of the identity.  Every such operator is similar to a normal operator. Now  quasi-Hermitian operators contain  spectral operators of scalar type with real spectrum  and, \emph{a fortiori},  self-adjoint operators. Thus we are led to study various generalized notions of (quasi-)similarity of operators, in particular in the unbounded case.

An additional motivation for such an analysis stems from recent developments in
the so-called Pseudo-Hermitian quantum mechanics (QM).
 This is an unconventional approach to QM, based on the use of a non-self-adjoint Hamiltonian, that can be transformed into  a self-adjoint one  by changing the ambient \hs, via a metric operator, as explained above.
  These Hamiltonians are in general assumed to be $\P\T$-symmetric, that is, invariant under the joint action of space reflection ($\P$) and complex conjugation ($\T$), and  they have often a real spectrum, usually discrete. In fact, they are quasi-Hermitian  or pseudo-Hermitian operators, the latter being slightly more general (see below).  A full analysis of $\P\T$-symmetric Hamiltonians may be found in the review paper of Bender \cite{bender}. 
  Since then, a large body of literature has been devoted to this topic.  An overview of the most recent works, including the various physical applications,  is presented    in a recent review  paper \cite{bender-specissue}.

The note is organized as follows.
We start by examining the relationship between metric operators and similarity, including a comparison with various concepts introduced in the literature. Then, in Section \ref{sect_quasisim}, we recall the notion of quasi-similarity, which is based on a bounded intertwining operator with unbounded inverse, then we extend it to the case of an unbounded intertwining operator.
Section \ref{sect_quasiH} is the heart of the paper, in which we examine systematically quasi-Hermitian operators in the case of a bounded  metric operator with possibly unbounded inverse.
Finally, in Section \ref{sec:unbddmetropappl}, we make contact with  unbounded metric operators and the application to quasi-Hermitian Hamiltonians.
\medskip

Before proceeding, we fix our notations and give some preliminary remarks. The framework is a separable \hs\ $\H$, with inner product
$\ip{\cdot}\cdot$, linear in the first entry. Then, for any operator $A$ in   $\H$, we denote its domain by $D(A)$, assumed to be dense in $\H$.

We start by analyzing the notion of metric operator  and the structure it generates in the ambient \hs.
\bedefi By a metric operator in a \hs\ $\H$, we mean a strictly
positive self-adjoint operator $G$, that is, $G>0$ or $\ip{G\xi}{\xi}\geq 0$ for every $\xi \in D(G)$
and $\ip{G\xi}{\xi}= 0$ if and only if $\xi=0$.
\findefi
Clearly, such an operator $G$ is densely defined and invertible, but need not be bounded; its inverse $G^{-1}$ is also a metric operator,  bounded or not (in this case,  $0\in \sigma_c(G)$, the continuous spectrum of $G$). We note that every $G^{\,\alpha} (\alpha \in\RN)$ is also a metric operator. We will use extensively $G^{\pm1/2}$
in the sequel.

If $G$ is a bounded metric operator, we denote by $\HG$ the Hilbert space obtained by completion of $\H$ with respect to the norm
$\norm{G}{\cdot}$ defined by the inner product
$$
 \ip{\xi}{\eta}_G:=\ip{G\xi}{\eta}= \ip{G^{1/2}\xi}{G^{1/2}\eta}, \quad \xi, \eta \in \H.
 $$
Thus $\norm{G}{\xi} =  \norm{}{G^{1/2}\xi}$ for every $\xi \in \H$.
The operator $G^{1/2}$ is an isometry of $(\H,\norm{G}{\cdot})$ into $(\H,\norm{}{\cdot})$.
In the same way, if $G^{-1}$ is unbounded,
the (dense) domain $D(G^{-1/2})$ may  be equipped with its graph norm, yielding the \hs\ $\H(G^{-1})$, densely embedded into $\H$.
Thus we obtain the following triplet:
\be \label{triplet}
 \H(G^{-1}) \;\subset\; \H \;\subset\;  \H(G),
\en
where all embeddings are continuous with dense range. Then $G^{-1/2}$ is unitary from $ \H(G^{-1})$ onto $\H$  and from $\H$ onto
$\H(G)$. Similarly, $G^{1/2}$ is unitary from $ \H(G)$ onto $\H$ and from $\H$ onto $\H(G^{-1})$.
If $G$ and  $G^{-1}$ are both bounded, the three spaces in the triplet \eqref{triplet} coincide as vector spaces and they carry different but equivalent norms.

\begin{figure}[t]
\centering \setlength{\unitlength}{0.5cm}
\begin{picture}(8,8)

\put(4.2,4){
\begin{picture}(8,8) \thicklines
 \put(-3.4,-0.9){\vector(3,1){2.2}}
\put(-3.4,2.2){\vector(3,1){2}}
 \put(-3.4,-1.9){\vector(3,-1){2.2}}
\put(-3.4,1.2){\vector(3,-1){2.2}}
\put(1.3,0.4){\vector(3,1){2.2}}
\put(1.3,-2.8){\vector(3,1){2.2}}
 \put(1.3,-0.4){\vector(3,-1){2.2}}
\put(1.45,2.6){\vector(3,-1){2.15}}
\put(0,3.2){\makebox(0,0){ $ \H(G^{-1})$}}
\put(-0.1,0){\makebox(0,0){ $\H$}}
\put(0,-3.2){\makebox(0,0){ $\H(G)$}}

\put(-5.3,1.5){\makebox(0,0){ $\H(R_{G^{-1}})$}}
\put(-5.2,-1.5){\makebox(0,0){$\H(R_{G})$}}
\put(5.3,1.5){\makebox(0,0){ $ \H(R_G^{-1})$}}
\put(5.3,-1.5){\makebox(0,0){$ \H(R_{G^{-1}}^{-1})$}}

\end{picture}
}
\end{picture}
\caption{\label{fig:diagram}The lattice of \hs s generated by a single metric operator (from Ref.\citen{pip-metric}, Fig. 1).}
 
\end{figure}
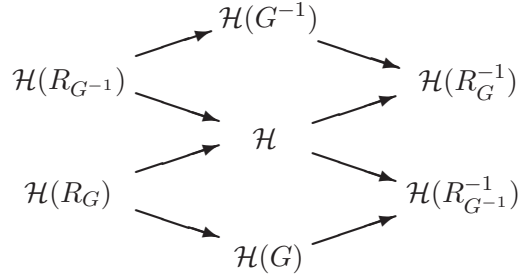

 On the other hand, if $G$ and  $G^{-1}$ are both unbounded, the three spaces in the triplet \eqref{triplet} are mutually not comparable. Instead they generate a lattice which has been thoroughly analyzed in  \cite[Section 2]{pip-metric}.
 First, consider the domain $D(G^{1/2})$. Equipped with its graph norm, this is a \hs, denoted $\H(R_G)$. That norm  may also be written as
 $$
\norm{R_G}{\xi}^2  = \ip{(1+G)\xi}{\xi} = \ip{R_G \xi}{\xi} =  \norm{ }{\xi}^2 + \norm{G}{\xi}^2 ,
$$
where $R_G = 1+G$ (which justifies the notation). Then define $\H(G)$ as the completion of   $\H(R_G)$ in the norm  $\norm{G}{\cdot}$, so that
$\H(R_G) = \H \cap \H(G)$, and  $\norm{R_G}{\cdot}$ is the projective norm, as used in interpolation theory \cite{berghlof }.
Next, the conjugate dual of $\H(R_G)$ is $ \H(R_G^{-1})= \H + \H(G^{-1})$, with  the inductive norm. Proceeding in the same way with $G^{-1}$, one gets the lattice of seven Hilbert spaces shown in Figure \ref{fig:diagram}, in which each arrow denotes a continuous map with dense range. Then the operator
$R_{G}^{1/2}$ is unitary from $\H(R_{G})$ onto $\H$, and from  $\H$ onto  $ \H(R_G^{-1})$.\footnote{The space $\H(R_G^{-1})$ is (three times) erroneously denoted $\H(R_{G^{-1}})$ in \cite[p.4]{pip-metric}  (see  Corrigendum). }
Hence $R_{G}$ is the Riesz unitary operator mapping $\H(R_{G})$ onto its conjugate dual  $ \H(R_G^{-1})$.
Similarly,  $R_{G^{-1}} $ is unitary from $\H(R_{G^{-1}})$ onto $ \H(R_{G^{-1}}^{-1})$.

At this point, it is necessary to clarify the relationship between metric operators and similar concepts commonly found in  the literature.
To start with, Dieudonn\'e \cite{dieudonne} calls \emph{quasi-Hermitian} a bounded operator $A$ on a \hs\ $\H$ for which there exists a { bounded},  strictly
positive,  self-adjoint operator $T\neq 0$ such that
\be\label{eq-quasiH}
TA=A \ha  T.
\en
Thus $T$ is invertible, but its inverse $T^{-1}$ need not be bounded. Such an operator $T$ is sometimes called \emph{quasi-invertible} \cite{hoover}.
Now, if the operator $T^{-1}$ is bounded, then \eqref{eq-quasiH} implies that $A$ is similar to a self-adjoint operator, thus it is a spectral operator of scalar type and real spectrum. However, the terminology about quasi-Hermitian operators  is not uniform in the literature. Kantorovitz \cite{kantor}  for instance,  defines quasi-Hermitian operators  exactly as spectral operators of scalar type and real spectrum. This is the case treated by Scholtz \emph{et al.} and Geyer \emph{et al.} \cite{scholtz, geyer} who introduced the concept in the physics literature.

A slightly more general notion is that of \emph{pseudo-Hermitian} operators, namely operators $A$ satisfying \eqref{eq-quasiH}, with $T$ and $T^{-1}$   bounded, but not necessarily positive {(this will unavoidably lead to indefinite metrics, see below).} This is the definition adopted also by Kretschmer-Szymanowski \cite{kretsch},  Mostafazadeh \cite{mosta1} or Albeverio \emph{et al.} \cite{alb-g-kuzhel}. 
Later on,  Kretschmer-Szymanowski, Mostafazadeh \cite{mosta2} and Bagarello-Znojil \cite{bag-zno2} adapted the definition to the case of an \emph{unbounded} operator $T$, claiming this is required for certain physical applications. Note that the last named authors have also coined the term \emph{cryptohermitian} for bounded quasi-Hermitian operators \cite{bag-zno2}.

Another issue to clarify is the relation between pseudo-Hermitian operators and $J$-self-adjoint operators in a Krein space.  Assume  that
$\H$ is a \hs\ with a \emph{fundamental symmetry} $J$, that is, a self-adjoint involution, $J=J\ha, J^2 = I$. Defining the projections $P_{\pm} = \frac12(I\pm J)$,
we obtain the fundamental decomposition of $\H$:
\be
\H = \H_+ \oplus  \H_-, \quad \H_{\pm} := P_{\pm}\H.
\en
Then, the space $\H$  endowed with the indefinite inner product
\be\label{indef-ip}
[\xi,\eta]_J := (J\xi, \eta)
\en
is a \emph{Krein space} \cite{bognar}.  Conversely, a Krein space may be defined as a decomposable, nondegenerate  inner product space $\K = \K_+ \oplus  \K_-$, with inner product  $[\cdot,\cdot]$, where the
subspace $\K_+$, resp. $ \K_-$, consists of vectors of positive, resp. negative norm $[\xi,\xi]^{1/2}$ and both subspaces $\K_{\pm}$ are complete in the
so-called intrinsic norm $|[\xi,\xi]|^{1/2}$. In that case, the $J$-inner product
\be
(\xi, \eta)_J := [J\xi,\eta]
\en
is positive definite and $\K$ is a \hs\ for the $J$-inner product.

Then,  a linear  densely defined operator $A$ in the Krein space $(\H, [\cdot,\cdot]_J )$ is called \emph{$J$-self-adjoint} if it satisfies the relation $A \ha  J = JA$.
Thus   $J$-self-adjoint operators  are pseudo-Hermitian and constitute the appropriate class to study rigorously, as claimed in \cite{alb-g-kuzhel,bognar}.

Finally, we note that non-self-adjoint operators (in Banach or \hs s) and their spectral properties are the object of a systematic analysis
by Davies \cite{davies}.

\section{SIMILAR AND QUASI-SIMILAR OPERATORS }
\label{sect_quasisim}

 In   this section we consider the notion of similarity of linear operators in Hilbert spaces and we focus in particular on the notion of quasi-similarity, as defined in our previous paper \cite{pip-metric} We divide the discussion into two parts. In the first one, the intertwining operator which determines the similarity or quasi-similarity is taken to be bounded; this assumption occurs often in the literature, but sometimes with very different meanings; so we take the occasion for comparing some of these definitions. In the second part of the section, we generalize some of the results obtained in \cite{pip-metric} to the case where the intertwining operator is not necessarily bounded.

From now on,
we will always suppose the domains of the given operators to be dense in $\H$.

\subsection{Quasi-similarity with bounded intertwining operators}
\label{sect_quasisim_2.1}

The notion of  similarity relies on that of intertwining operators, that we first take as bounded.

\bedefi \label{def:qu-sim}
Let $\H, \K$ be Hilbert spaces, $D(A)$ and $D(B)$ dense subspaces
of $\H$ and $\K$,  respectively,   $A:D(A) \to \H$, $B: D(B) \to \K$ two linear operators.
 A bounded operator $T:\H \to \K$ is called an \emph{intertwining operator}  for $A$ and $B$ if
\begin{itemize}
\vspace{-2mm}\item[({\sf io$_1$})] $T:D(A)\to D(B)$;
\vspace{-2mm}\item[({\sf io$_2$})]$BT\xi = TA\xi, \; \forall\, \xi \in D(A)$.
\end{itemize}
\findefi

\bedefi
Let  $A, B$ be two linear operators in the Hilbert spaces $\H$ and $\K$, respectively. Then
we say that $A$ and $B$ are \emph{similar}, and write $A\sim B$,
if there exists a (bounded) intertwining operator $T$ for $A$ and $B$ with bounded inverse $T^{-1}:\K\to \H$, which is intertwining for $B$ and $A$.
\findefi
We notice that  $\sim$ is an equivalence relation and that  one has $TD(A) = D(B)$. If $T$  is unitary,
$A$ and $B$ are unitarily equivalent, which we denote by $A\stackrel{u}{\sim} B$.

\smallskip

The notion of similarity  is too strong in many situations, thus we seek a weaker one. A natural step is to drop the boundedness of $T^{-1}$. Thus, following \citen{pip-metric}, we define:
\bedefi\label{def:quasi-sim}
We say that $A$  is \emph{quasi-similar} to   $B$, and write $A\dashv B$, if there exists an intertwining operator $T$ for $A$ and $B$ which is  invertible, with inverse $T^{-1}$   densely defined (but not necessarily bounded).
\findefi
Note that, even if $T^{-1}$ is bounded, $A$ and $B$ need not be similar, unless $T^{-1}$ is also  an intertwining operator. Also,  if $A\dashv B$, with intertwining operator $T$, then $B^*\dashv A^*$ with intertwining operator $T^*$.

\berem   There is a considerable confusion in the literature concerning the notion of quasi-similarity.

 {(1)  First, essentially all authors consider only (quasi-)similarity between two \emph{bounded} operators.
 Next, a bounded invertible operator $T$ with (possibly) unbounded inverse $T^{-1}$ is called a \emph{quasi-affinity} by Sz.-Nagy and  Foia\c{s} \cite[Chap.II, Sec.3]{nagy} and  a  \emph{quasi-invertible} operator by other authors \cite{hoover}. Then, if  $A,B$ are two  bounded operators such that $TA=BT$, that is, $A\dashv B$,   $A$ is called a \emph{quasi-affine transform} of $B$. In this context, $A$ and $B$  are called \emph{quasi-similar}  if $A\dashv B$ \emph{and} $B\dashv A$  (so that quasi-similarity becomes also an equivalence relation). }

 (2) Tzafriri \cite{tzafriri} considers only bounded spectral operators,
 in Dunford's sense \cite{dunford, dunford-schwartz}.  For these, he introduces a different notion of quasi-similarity (but under the same name) based on the resolution of the identity.

Hoover \cite{hoover} shows that if two  bounded spectral operators $A$ and $B$ are quasi-similar (i.e. $A\dashv B$ and $B\dashv A$), then $B$ is quasi-similar
to $A$ in the sense of Tzafriri (which he calls weakly similar).  On the other hand, Feldzamen \cite{feldzamen} considers yet another notion of generalized similarity, called \emph{semi-similarity}, but then Hoover shows that two  semi-similar bounded spectral operators $A$ and $B$ are in fact quasi-similar.

(3)  Quasi-similarity of unbounded operators  is considered by \^{O}ta and  Schm\"udgen \cite{ota-schm}.   Namely, given two unbounded operators
$A$ and $B$ in \hs s $\H , \K$, respectively,  $A$ is said to be \emph{quasi-similar} to  $B$ if there exist two (quasi-invertible) intertwining operators (in the sense of Definition \ref{def:qu-sim}) $T_{AB}: D(A)\to D(B)$ and $T_{BA}: D(B)\to D(A)$. In other words, this notion is the straightforward generalization of the quasi-similarity of bounded operators defined by the previous authors.
\enrem

 {In the sequel, we will stick to the asymmetrical notion of quasi-similarity given in Definition
\ref{def:quasi-sim}, namely, $A\dashv B$. The reason is that, in physical applications, one expects the operator $B$, acting in the different \hs\ $\K$, to be   better behaved than $A$ acting in $\H$, so that the relation between the two should not always be symmetrical.

Accordingly,  we will say that
two closed operators $A$ and $B$ are \emph{mutually quasi-similar} if they are quasi-similar in the sense of
\^{O}ta and  Schm\"udgen, that is, if we have both  $A\dashv B$ and  $B\dashv A$, which we denote by
$A\dashv \vdash B$. Clearly $\dashv \vdash $ is an equivalence relation. Moreover, $A\dashv \vdash B$ implies
$A\ha \dashv \vdash B\ha$.}
\smallskip

 Before leaving the section, it is worth quoting a result from  \^{O}ta and  Schm\"udgen \cite{ota-schm}. 
 \beprop
 {Let $A$ and $B$ be closed operators. Then:
\begin{itemize}
\vspace{-2mm}\item[(i)] Let $A$ and $B$ be  normal (in particular, self-adjoint) and $A\dashv \vdash B$. Then
they are unitarily equivalent, $A\stackrel{u}{\sim} B$.
\vspace{-2mm}\item[(ii)] Let $A$ be symmetric and $B$ self-adjoint, with $A\dashv \vdash B$. Then $A$ is self-adjoint
and $A\stackrel{u}{\sim} B$.
\vspace{-2mm}\item[(iii)] Let $A$ be symmetric  and $A\dashv \vdash A\ha$. Then $A$ is self-adjoint.
\eni
}\enprop

\subsection{Quasi-similarity with unbounded intertwining operators}
\label{sect_quasisim_2.2}
Now it is an easy step to generalize the preceding analysis to the case of an \emph{unbounded} intertwining operator.
Namely we define
\bedefi \label{def_36}
Let $A,B$ two  densely defined  linear operators on the Hilbert spaces  $\H, \K$, respectively.
 A closed (densely defined) operator $T:\H \to \K$ is called an \emph{intertwining operator}  for $A$ and $B$ if
\begin{itemize}
\vspace{-2mm}\item[({\sf io$_0$})] $D(TA) = D(A)\subset D(T)$;
\vspace{-2mm}\item[({\sf io$_1$})] $T:D(A)\to D(B)$;
\vspace{-2mm}\item[({\sf io$_2$})]$BT\xi = TA\xi, \; \forall\, \xi \in D(A)$.
\end{itemize}
\findefi
The first part of condition ({\sf io$_0$}) means that $\xi\in D(A)$ implies $A\xi\in D(T)$.
Of course, this definition reduces to the usual one, Definition \ref{def:qu-sim}, if $T$ is bounded, since then condition ({\sf io$_0$})
is satisfied automatically.

 In terms of this definition, we say again
 that $A$ is \emph{quasi-similar} to   $B$, and write $A\dashv B$, if there exists a (possibly unbounded) intertwining operator $T$ for $A$ and $B$ which is  invertible, with inverse $T^{-1}$   densely defined.

 {From the previous definition it follows easily that $A$ is quasi-similar to $B$ if, and only if, $A \subseteq T^{-1}BT$, where $T$ is a closed densely defined operator which is injective and has dense range.

Quasi-similarity with an unbounded intertwining operator, however, may occur only under {\em singular} circumstances.

\begin{prop} Let $A\dashv B$ with intertwining operator $T$. If  the resolvent set $\rho(A)\ne \emptyset$, then $T$ is necessarily bounded.

\end{prop}
 {\bf Proof. } From $A \subseteq T^{-1}BT$ it follows that $A-\lambda I \subseteq T^{-1}(B-\lambda I)T$, for every $\lambda \in {\mb C}$. If $\lambda \in \rho(A)$, then, for every $\eta \in \H$, there exists $\xi \in D(A)$ such that $(A-\lambda I)\xi =\eta$. Thus, $\xi \in D(T^{-1}(B-\lambda I)T)$ and
$T^{-1}(B-\lambda I)T\xi=\eta$. This clearly implies that $\eta \in D(T)$. Hence $D(T)=\H$ and $T$ is bounded.
\qed

\medskip
}
In our previous paper \cite{pip-metric},   
we have analyzed the relationship between the spectral properties of two similar or quasi-similar operators. First,     similarity preserves the resolvent set $\rho(\cdot)$ of operators and the different parts of the spectrum, namely, the point spectrum $\sigma_p (\cdot)$, the continuous spectrum $\sigma_c(\cdot)$ and the residual spectrum $\sigma_r(\cdot)$.
Notice that, here as in   \cite{pip-metric}, we follow Dunford-Schwartz \cite{dunford-schwartz},  whose definition implies that
the three sets $\sigma_p (A), \sigma_c(A), \sigma_r(A)$ are disjoint and
\be\label{eq:spec}
\sigma (A) = \sigma_p (A) \cup \sigma_c(A)\cup \sigma_r(A).
\en
However, other authors give a different definition of the continuous spectrum, implying that it is no longer disjoint from the point spectrum, for instance, Reed-Simon \cite{reedsim1} or Schm\"{u}dgen \cite{schm}.  This alternative definition allows for eigenvalues embedded in the continuous spectrum, a situation common in many physical situations, such as the Helium atom, and a typical source of   resonance effects in scattering theory
 (see \cite[Sect. XII.6]{reedsim4}).

Next, in Propositions 3.24--3.28 of   \cite{pip-metric}, we have given a series of results concerning the relationship
 between the spectral properties of two quasi-similar operators, in the usual case of a bounded intertwining operator.
Actually some of these remain true when the intertwining operator $T$ is unbounded, as we now show. The first result parallels part of Proposition 3.24 of 
 \cite{pip-metric}.

\begin{prop}\label{prop_sigmap-unbdd} Let $A$ and $B$ be closed operators and assume that $A\dashv B$,  with the (possibly unbounded) intertwining operator $T$. Then the following statements hold.
\begin{itemize}
\vspace{-2mm}\item[(i)]
$\sigma_p(A)\subseteq \sigma_p(B)$: If $\xi \in D(A)$ is an eigenvector of $A$ corresponding to the eigenvalue $\lambda$,
 then $T\xi$ is an eigenvector of $B$ corresponding to the same eigenvalue.
Thus, for every $\lambda \in \sigma_p(A)$, one has $m_A(\lambda) \leq m_B(\lambda)$, {where $m_C(\lambda)$ denotes the multiplicity of $\lambda$ as eigenvector of $C\in\{A, B\}$}.
\vspace{-2mm}\item[(ii)] If $TD(A) = D(B)$ and $T^{-1}$ is bounded, then $\sigma_p(A)= \sigma_p(B)$.
\vspace{-2mm}\item[(iii)] If $T^{-1}$ is bounded and $TD(A)$ is a core for B,  then $\sigma_p(B)\subseteq \sigma(A)$.
\end{itemize}
\end{prop}
 {\bf Proof. }
 (i) Let $\lambda \in \sigma_p(A)$ :  there is $\psi\in D(A), \psi\neq 0$, such that $ A \psi = \lambda \psi$. Then, by ({\sf io$_0$}), $A \psi\in D(T)$ and
$ TA \psi = \lambda T\psi$. The rest is obvious.

(ii) If $\eta\in D(B)$, there exists $\xi\in D(A)$ such that $\eta = T\xi$ and $T^{-1}\eta =  \xi$. Then
$B\eta=  \lambda\eta$  implies $A\xi=  \lambda\xi$.

(iii) Let $\lambda \in \sigma_p(B)$. Then there exists $\eta \in D(B)\setminus\{0\}$ such that $B\eta= \lambda \eta$.
We may assume that $\|\eta\|=1$.
Since  $T(D(A))$ is a core for $B$, there exists a sequence $\{\xi_n\}\subset D(A)$ such that $T\xi_n \to \eta$ and $BT\xi_n \to B\eta$. Then,
$$
\lim_{n \to \infty} T(A\xi_n -\lambda \xi_n) = \lim_{n \to \infty} TA\xi_n -\lambda \lim_{n \to \infty}T\xi_n=\lim_{n \to \infty}BT\xi_n -\lambda \eta
= B\eta -\lambda \eta =0.
$$
Since $T^{-1}$ is bounded, we have  $(A\xi_n -\lambda \xi_n) \to 0.$
Assume that $\lambda \in \rho(A)$, so that $(A-\lambda I)^{-1} \in \BH$. If we define  $\eta_{n}= (A-\lambda I)\xi_{n}$, we get
 $\eta_{n} \to 0$. Hence, $\xi_{n}= (A-\lambda I)^{-1}\eta_{n}\to 0$.
Since $T\xi_{n} \to \eta$ and $T$ is closed, this   implies
that $T\xi_{n} \to 0$,   which is impossible since $\|\eta\|=1$.
\qed
\medskip

Notice that we cannot say anything about residual spectra, because these are defined on the basis of the relation  $\sigma_r(A)= \ov{\sigma_p(A \ha )}  =
\{ \ov{\lambda}: \lambda \in \sigma_p(A \ha )\}$ and $A\dashv B$ does \emph{not} imply $B\ha \dashv A\ha$ in general for an unbounded intertwining operator. Thus we can only state (compare   \cite[Corollary 3.26]{pip-metric}):
\becor \label{cor_3.24-unbdd} Let $A$, $B$ be as in Proposition \ref{prop_sigmap-unbdd} and assume that $T^{-1}$ is {bounded and everywhere defined}. Then
$$
\rho(A)\setminus \sigma_p(B) \subseteq \rho(B).
$$
\encor

Things improve if we assume the operators $A$ and $B$ to be mutually quasi-similar,
\mbox{$A\dashv \vdash B$.}
First, we can improve Proposition \ref{prop_sigmap-unbdd}.
\begin{prop}\label{prop_mutsigmap}
 Let $A$ and $B$ be closed operators and assume that $A\dashv \vdash B$,
with  possibly unbounded intertwining operators  $T_{AB}: D(A)\to D(B)$ and $T_{BA}: D(B)\to D(A)$. Then:
\begin{itemize}
\vspace{-2mm}\item[(i)] $\sigma_p(A)=\sigma_p(B)$:
if $\xi \in D(A)$ is an eigenvector of $A$ corresponding to the eigenvalue $\lambda$,
 then $T_{AB}\xi$ is an eigenvector of $B$ corresponding to the same eigenvalue;
 if $\eta \in D(B)$ is an eigenvector of $B$ corresponding to the eigenvalue $\mu$,
 then $T_{BA}\xi$ is an eigenvector of $A$ corresponding to the same eigenvalue. In both cases, the multiplicities are the same.
\vspace{-2mm}\item[(ii)] Assume that both intertwining operators $T_{AB},T_{BA}$ have a bounded inverse.
Then one has, in addition, $\rho(A) = \rho(B)$, thus also $\sigma (A)= \sigma (B)$.
 \eni
 \end{prop}
 {\bf Proof. }
 (i) is obvious.

(ii) Let $T_{AB}^{-1}$ be everywhere defined and bounded. Then, by Corollary \ref{cor_3.24-unbdd} and (i),
we have
$$
\rho(A)\setminus \sigma_p(B) = \rho(A)\setminus \sigma_p(A) = \rho(A)\subseteq \rho(B).
$$
Exchanging $A$ and $B$, we get  $\rho(B)\subseteq \rho(A)$, which proves (ii).
\qed

Under these conditions, it follows that $\sigma_c (A) \cup \sigma_r (A) = \sigma_c (B) \cup \sigma_r (B)$, but
we cannot say more, for the same reason as before.

\section{QUASI-HERMITIAN AND QUASI-SELF-ADJOINT OPERATORS}
\label{sect_quasiH}

Intuitively, a quasi-Hermitian operator $A$ is an operator which is Hermitian when the space is endowed with a new inner product.
We will make this precise in the sequel, generalizing the original definition of Dieudonn\'e \cite{dieudonne}. 

\bedefi \label{quasihermitian}
A closed operator $A$, with dense domain $D(A)$ is called \emph{quasi-Hermitian} if there exists a metric operator $G$, with dense domain $D(G)$, such that $D(A)\subset D(G)$ and
\be \label{eq_quasihermitian}
\ip{A\xi}{G\eta}= \ip{G\xi}{A\eta}, \quad \xi, \eta \in  D(A)\en
\findefi
Of course, if the condition $D(A)\subset D(G)$ is not satisfied,   the relation  \eqref{eq_quasihermitian} may hold for every
$ \xi, \eta \in D(G)\cap D(A)$, but
the definition could be meaningless, since it may happen that $D(G)\cap D(A)=\{0\}$. For this reason, this more general definition would require additional conditions on $G$, that will, of course, depend whether $G$ and $G^{-1}$ are bounded or not.

\subsection{Changing the metric}

Take first $G$   bounded and $G^{-1}$ possibly unbounded. According to the analysis of Section \ref{sect_intro}, we are facing the triplet \eqref{triplet},
where $\H(G)$ is a    \hs\, equipped with the norm $\norm{G}{\cdot}$.
 This \hs\ is obtained from $\H$ by changing the metric, thus the question is how operator properties are transferred from $\H$  to $\H(G)$.
In particular,  several adjoints may be defined and we have to compare them before analyzing quasi-Hermitian operators.

We will often use the following well-known result (see, e.g.,  \cite[Theor. 4.1]{weidmann}).
\begin{lem} \label{lem_25} Let us consider a closed operator $S$ in $\H$ with dense domain $D(S)$.
 If $B$ is bounded, one has $ (BS) \ha =S \ha B\ha$.
\end{lem}

Let again   $S$  be a closed densely defined operator in $\H$. {Then  $D(S)$ is dense in $\HG$ (see  \cite[Lemma 3.1]{pip-metric})}.
So $S$ has a well-defined adjoint in $\HG$. We denote it by $S^\#$, while we denote, as usual,  the adjoint in $\H$  by $S \ha $. Let us compute $S^\#$.

\beprop \label{prop_242} Let $G$ be a bounded metric operator in $\H$ and $S$ a closed, densely defined operator in $\H$. Then:

 (i)  $\widetilde{G}D(S^\#) \subseteq D(S \ha )$ and $S \ha \widetilde{G}\eta= \widetilde{G}S^\#\eta$, for every $\eta\in D(S^\#)$, where $\widetilde{G}$ denotes the natural extension of $G$ to $\HG$.

(ii) If $G^{-1}$ is also bounded, then   $D(S^\#)=G^{-1}D(S \ha )$ and $S^\#\eta= G^{-1}S \ha G\eta$, for every $\eta\in D(S^\#)$.
\enprop
\bdim  (i) Let $\eta \in D(S^\#)$. Then there exists $\eta^\#\in \HG$ such that
$$
 \ip{S\xi}{\eta}_G= \ip{\xi}{\eta^\#}_G, \quad \forall \xi\in D(S)
$$
and $S^\#\eta=\eta^\#$.
Since
$$
 |\ip{S\xi}{\eta}_G|= |\ip{\xi}{\eta^\#}_G|  \leq  \|\xi\|_G \, \|\eta^\#\|_G  \leq   \gamma \|\xi\| \, \|\eta^\#\|_G, \quad \forall \,\xi\in D(S),
 $$
there exists $\eta \ha \in \H$ such that
$$
 \ip{S\xi}{\widetilde{G}\eta}=\ip{\xi}{\eta \ha }.
 $$
 This implies that $\widetilde{G}\eta\in D(S \ha ).$ We recall that $\widetilde{G}\HG = \H(G^{-1}) \subset  \H$.
It is easily seen that $\eta \ha = \widetilde{G}\eta^\#$.
Hence
 $$
 \ip{S\xi}{\widetilde{G}\eta}=  \ip{\xi}{S \ha \widetilde{G}\eta }=\ip{\xi}{\widetilde{G}S^\#\eta}, \quad \forall\, \xi \in D(S).
 $$
This, in turn, implies that $S \ha \widetilde{G}\eta= \widetilde{G}S^\#\eta$, for every $\eta\in D(S^\#)$.

(ii) If $G$ and $G^{-1}$ are both bounded, $\H(G^{-1}) = \H = \H(G)$ as vector spaces, but with different norms.
From (i), it follows that $GD(S^\#) \subseteq D(S\ha)$ and $S \ha {G}\eta= {G}S^\#\eta$, for every $\eta\in D(S^\#)$.

Let  now $\zeta\in D(S\ha)$. Then $\zeta=G\eta$ and $S\ha G \eta = G \eta\ha$, for some $\eta, \eta\ha\in \H$.
Then we have, for any $\xi\in D(S)$,
$$
\ip{S\xi}{\eta}_G = \ip{S\xi}{G\eta} = \ip{S\xi}{\zeta}= \ip{\xi}{S\ha \zeta} =  \ip{\xi}{G\eta\ha} =  \ip{\xi}{\eta\ha}_G.
$$
Hence $\eta\in D(S^\#)$ and $S^\#\eta = \eta\ha = G^{-1}S\ha \zeta=  G^{-1}S\ha G \eta$, i.e., $GS^\#\eta = S\ha G \eta$.
\edim
\medskip

The second part of the proof of (ii) does not hold if $G^{-1}$ is unbounded. In this case, we have only the following partial result.
\beprop
Let $G$ be a metric operator with $G^{-1}$ possibly unbounded. Then, for every $\zeta\in D(S\ha) \cap D(G^{-1/2})$ such that $S\ha \zeta \in D(G^{-1/2})$,
there exists $\eta\in  D(S^\#)$ such that $\widetilde{G} \eta = \zeta$ and $GS^\# \eta = S\ha \widetilde{G} \eta$.
\enprop
\bdim
The proof is similar to the previous one,
noting that $D(G^{-1/2}) = \H(G^{-1})$ and $\zeta, S\ha \zeta \in \H(G^{-1})$. Hence there exist $\eta, \eta\ha \in \H$ such that
$$
\widetilde{G^{-1/2}}\zeta=\widetilde{G^{1/2}}\eta \quad \mbox{and} \quad  \widetilde{G^{-1/2}}\zeta \ha = \widetilde{G^{1/2}}\eta \ha .
$$
 The rest is as before.
\edim

\belem \label{lemma_onehalf2}Let $G$   be   bounded in $\H$ and let $S$ be closed and densely defined.
Define $D(K)=G^{1/2}D(S)$  and  $K\xi={G^{1/2}}SG^{-1/2}\xi$, $\xi \in D(K)$. Then $K$ is densely defined and $K \ha ={G^{-1/2}}S \ha G^{1/2}.$
\enlem
\bdim It is easy to see that $D(K)$ is dense in $\H$ and that ${G^{-1/2}}S \ha G^{1/2}\subset K \ha $. We prove the converse inclusion.
Let $\eta \in D(K \ha )$. Then there exists $\eta \ha  \in \H$ such that
$$
\ip{{G^{1/2}}SG^{-1/2}\xi}{ \eta}= \ip{\xi}{\eta \ha }, \quad \forall\, \xi \in G^{1/2}D(S).
$$
This implies that
$$
 \ip{S\zeta}{{G^{1/2}} \eta}= \ip{G^{1/2}\zeta}{\eta \ha }=\ip{\zeta}{G^{1/2}\eta \ha }, \quad \forall\, \zeta \in D(S).
 $$
Hence, ${G^{1/2}} \eta \in D(S \ha )$ and
$$
 \ip{G^{-1/2}\xi}{{S \ha G^{1/2}} \eta}= \ip{\xi}{\eta \ha }, \quad \forall\, \xi\in  G^{1/2}D(S).
 $$
This in turn implies that $S \ha G^{1/2} \eta\in D(G^{-1/2})$ and $K \ha \eta={G^{-1/2}}S \ha G^{1/2}\eta$.
\edim

In Proposition \ref{prop_242} (ii), we have obtained the expression of $S^\#$ in the case where $G$ and $G^{-1}$ are both bounded in $\H$, namely, $S^\#=G^{-1}S \ha G $. If $G^{-1}$ is unbounded, we can  characterize  the restriction to $\H$ of $S^\#$ (see   \cite [Prop. 3.19]{pip-metric}.
\begin{prop}\label{prop_214}Given  the closed  operator  $S$, put
\be\label{eq-domain}
 D(S^{\,\sharp}):= \{\eta \in \H: G\eta \in D(S \ha ), \, S \ha  G\eta \in D(G^{-1}) \}
\en
and
$$
S^{\,\sharp} \eta := G^{-1}S \ha  G\eta, \quad \forall\, \eta \in D(S^{\,\sharp}).
$$
Then $S^{\,\sharp}$ is the restriction to $\H$ of the adjoint $S^\#$ of $S$ in $\H(G)$.
\end{prop}

\becor \label{cor25}
If the domain $ D(S^{\,\sharp})$ is dense, then $S^\#$ is densely defined and $S$ is closable in $\H(G)$.
\encor
However, we still don't know whether $(S^\#)^\# = S$, i.e., whether $S$ is  closed in $\H(G)$.
\medskip

\medskip

An interesting application of the operator $S^{\,\sharp}$ is the following \cite[Lemma 3.23]{pip-metric}.
\belem Let $A,B$ be closed and  $A\dashv B$ with a bounded metric intertwining operator $G$. Then
$A^{\,\sharp}$ is densely defined, $B_0:= (A^{\,\sharp}) \ha $  is minimal among the closed operators $B$ satisfying, for fixed $A$ and $G$, the conditions
\begin{align*} & G:D(A)\to D(B);\\
& BG\xi = GA\xi, \; \forall\, \xi \in D(A),
\end{align*}
i.e., $B_0$ is minimal among the closed operators $B$ satisfying $A\dashv B$.
Moreover,  $GD(A)$ is a core for $B_0$.
\enlem

\subsection{Bounded quasi-Hermitian operators}

Let $A$ be a bounded operator in $\H$. Assume that $A$ is quasi-Hermitian in the sense of Definition \ref{quasihermitian}
 and that the metric operator $G$  is bounded with  bounded inverse. Then
\be
\ip{GA\xi}{\eta}=\ip{A\xi}{G\eta} = \ip{G\xi}{A\eta}= \ip{\xi}{GA\eta}, \quad \forall\, \xi, \eta \in \H.
\en
This implies that $GA$ is self-adjoint.
\beprop Let $A$ be bounded.
The following statements are equivalent.
\begin{itemize}
\vspace*{-2mm}\item[(i)] $A$ is quasi-Hermitian.
\vspace*{-2mm}\item[(ii)] There exists a bounded metric operator $G$, with bounded inverse, such that $GA (=A \ha G)$ is self-adjoint.
\vspace*{-2mm}\item[(iii)]{ $A$ is similar to a self-adjoint operator $K$ and the corresponding intertwining operator is metric}.
\end{itemize}
\enprop
\bdim \s{(i)$\Rightarrow$(ii)} is easy.
\\[1mm]
\s{(ii)$\Rightarrow$(iii}):
We put $K=G^{1/2}AG^{-1/2}$. Since $A \ha G$ is self-adjoint we get
$$K \ha =G^{-1/2}A \ha G^{1/2}=G^{-1/2}(A \ha G)G^{-1/2}=G^{-1/2}(GA)G^{-1/2}=G^{1/2}AG^{-1/2}$$
Hence, $K$ is self-adjoint and $A=G^{-1/2}KG^{1/2}$, i.e., $A\sim K$.
\\[1mm]
\s{(iii)$\Rightarrow$(i)}:
Assume $A=G^{-1/2}KG^{1/2}$ with $K=K \ha $. Then, for every $\xi, \eta \in \H$
\begin{align*}
\ip{GA\xi}{\eta}&=\ip{A\xi}{G\eta}=\ip{G^{-1/2}KG^{1/2}\xi}{G\eta}
\\&=\ip{G^{1/2}\xi}{KG^{1/2}\eta}=\ip{G^{-1/2}G\xi}{KG^{-1/2}G\eta}
\\&=\ip{G\xi}{G^{-1/2}KG^{1/2}\eta}=\ip{G\xi}{A\eta}.
 \\[-14mm]
&\end{align*}
\qed

\noi Thus $A$ is  a spectral operator  of scalar type and real spectrum \cite{dunford,dunford-schwartz}. 

\subsection{Unbounded quasi-Hermitian operators}

Let again $G$   be bounded, but now we take $A$ unbounded and quasi-Hermitian in the sense of Definition \ref{quasihermitian}.

\beprop \label{prop410}
If $G$ is
bounded, then  $A$ is quasi-Hermitian if, and only if, $GA$ is symmetric in $\H$.
\enprop
\bdim If $A$ is quasi-Hermitian, from \eqref{eq_quasihermitian}, it follows immediately that
\be
\ip{GA\xi}{\eta}=\ip{A\xi}{G\eta} = \ip{G\xi}{A\eta}= \ip{\xi}{GA\eta}, \quad \forall\, \xi, \eta \in D(A).
\en
Hence $GA$ is symmetric.

On the other hand, if $GA$ is symmetric,
\be
\ip{A\xi}{G\eta} =\ip{GA\xi}{\eta}= \ip{\xi}{GA\eta} = \ip{G\xi}{A\eta} , \quad \forall\, \xi, \eta \in D(A).
\en
Thus, $A$ is quasi-Hermitian.
\edim
\medskip

The next step consists in investigating the self-adjointness of $A$ as an operator in $\HG$.

\beprop \label{prop_28}Let  $G$ be bounded. If $A$ is self-adjoint in $\HG$, then $GA$ is symmetric in $\H$.
If  $G^{-1}$ is also bounded, then  $A$ is self-adjoint in $\HG$ if, and only if, $GA$ is self-adjoint  in $\H$.
\enprop
\bdim Let $A= A^\#$. Then, by Lemma \ref{lem_25}, $(GA) \ha =A \ha G, \, \forall\, \xi\in D(A)$,
$GD(A) \subseteq D(A \ha )$ and $A \ha G\xi =GA\xi$, by Proposition \ref{prop_242}(i).
Hence $GA$ is symmetric.

If  $G^{-1}$ is bounded, one has, by Proposition \ref{prop_242}(ii),
\smallskip

\centerline{$A=A^\#= G^{-1}A \ha G \Leftrightarrow GA=A \ha G=(GA) \ha  \Leftrightarrow GA $   self-adjoint . }

 \vspace*{-5mm}
 \qed

Now we turn the problem around.  Namely, given the closed densely defined operator $A$, possibly unbounded, we seek whether there is a metric operator $G$ that makes $A$ quasi-Hermitian  and self-adjoint in $\HG$. The first result is rather strong.

\beprop \label{prop_29}  {Let $A$ be  closed and densely defined}. Then the following statements are equivalent:
\begin{itemize}
\vspace*{-2mm}\item[(i)]There exists a bounded metric operator $G$, with bounded inverse, such that $A$ is self-adjoint in  $\H(G)$.
\vspace*{-2mm}\item[(ii)]There exists a bounded metric operator $G$, with bounded inverse, such that $GA=A \ha G$, i.e.,  $A$ is similar to its adjoint $A \ha $, with intertwining operator $G$.
\vspace*{-2mm}\item[(iii)]There exists a bounded metric operator $G$, with bounded inverse, such that $G^{1/2} A G^{-1/2}$ is self-adjoint.
\vspace*{-2mm}\item[(iv)] $A$ is a spectral operator of scalar type with real spectrum.

\end{itemize}
\enprop
\bdim (i) $\Rightarrow$ (ii)  is clear.  We   prove that (ii) $\Rightarrow$ (iii)  $\Rightarrow $ (i).
\\[1mm]
\s{(ii) $\Rightarrow$ (iii)}: Define $K=G^{1/2}AG^{-1/2}$. We prove that $K=K \ha $. Indeed, by Lemma \ref{lemma_onehalf2}, we have
$$
 K \ha = G^{-1/2}A \ha G^{1/2}=G^{-1/2}A \ha GG^{-1/2}= G^{-1/2}GAG^{-1/2}=G^{1/2}AG^{-1/2}=K.
 $$
\s{(iii) $\Rightarrow$ (i)}: Let $G^{1/2} A G^{-1/2}$ be self-adjoint. Then by Lemma \ref{lemma_onehalf2}, we get
$$
GA= G^{1/2}(G^{1/2}AG^{-1/2})G^{1/2}= G^{1/2}(G^{-1/2}A \ha G^{1/2}) G^{1/2}= A \ha G= (GA) \ha .
$$
The statement follows from Proposition \ref{prop_28}.
\\[1mm]
\s{(iii) $\Rightarrow$ (iv)}: Put $H = G^{1/2} A G^{-1/2}$, then $H$ is self-adjoint. Hence $H=\int_\RN \lambda \ud E(\lambda)$, where $\{E(\lambda)\}$
is a self-adjoint spectral family. From this it follows that
$$
A=\int_\RN \lambda \ud X(\lambda), \quad \mbox{where} \quad X(\lambda)= G^{-1/2}E(\lambda)G^{1/2}.
$$
Hence, $A$ is a spectral operator of scalar type. Moreover, as seen above, $\sigma(A) = \sigma(H) \subseteq \RN$.
\\[1mm]
\s{(iv) $\Rightarrow$ (iii)}:  If $A$ is a spectral operator of scalar type with $\sigma(A)  \subseteq \RN$, then
$A=\int_\RN \lambda \ud X(\lambda)$, where $\{X(\lambda)\}$  is a countably additive resolution of the identity \cite{inoue-trap}. 
By a result of Mackey   \cite[Theor. 55]{mackey},    there exists a bounded operator $T$ with bounded inverse and a self-adjoint resolution of the identity $\{E(\lambda)\}$
such that  $X(\lambda) = T^{-1} E(\lambda) T$. Put $G = |ÊT |^2$. By the polar decomposition, $T = U G^{1/2}$ with $U$ unitary. Hence,
$X(\lambda) = G^{-1/2} U^{-1} E(\lambda) U G^{1/2}$. Put $F(\lambda) = U^{-1} E(\lambda) U$. Then $\{F(\lambda)\}$ is a self-adjoint resolution of the identity. Thus $H:=\int_\RN \lambda \ud F(\lambda)$ is self-adjoint. Clearly $H = G^{1/2} A G^{-1/2}$, as announced.\footnote{
Using Dunford's result  \cite[Sect. XV.6]{dunford-schwartz},  we may conclude directly that $A = T^{-1} S T$, with $S$ self-adjoint. The rest follows by putting again $T = U G^{1/2}$.}
\edim

  Condition (i) of Proposition \ref{prop_29} suggests the following definition, slightly more general, in that we do not require $G^{-1}$ to be bounded.

\bedefi Let $A$ be closed and densely defined. We say that $A$ is \emph{quasi-self-adjoint} { if there exists a bounded metric operator} $G$, 
such that $A$ is self-adjoint in  $\H(G)$.\findefi

In particular, if any of the conditions of Proposition \ref{prop_29} is satisfied, then $A$ is quasi-self-adjoint.

 Actually, Proposition \ref{prop_29} characterizes quasi-self-adjointness in terms of similarity of $A$ and $A^*$, if the intertwining metric operator is bounded with bounded inverse.
Instead of requiring that $A$ be similar to $A \ha $,   we may ask  that they be only quasi-similar.   The price to pay is that now $G^{-1}$ is no longer bounded and, therefore, the equivalences stated in Proposition \ref{prop_29} are no longer true.

\beprop \label{prop_292}  {Let $A$ be  closed and densely defined}.  Consider the statements
\begin{itemize}

\vspace*{-2mm}\item[(i)]There exists  a bounded metric operator $G$ such that $GD(A)= D(A \ha )$, $A \ha G\xi =GA\xi$, for every $\xi \in D(A)$, in particular, $A$ is quasi-similar to its adjoint $A \ha $, with intertwining operator $G$.

\vspace*{-2mm}\item[(ii)]There exists a bounded metric operator $G$, such that $G^{1/2} A G^{-1/2}$ is self-adjoint.
\vspace*{-2mm}\item[(iii)]There exists a bounded metric operator $G$  such that $A$ is self-adjoint in $\HG$; { i.e., $A$ is quasi-selfadjoint}.
\vspace*{-2mm}\item[(iv)]{ There exists a bounded metric operator $G$  such that $GD(A)= D(G^{-1}A \ha )$, $A \ha G\xi =GA\xi$, for every $\xi \in D(A)$, in particular, $A$ is quasi-similar to its adjoint $A \ha $, with intertwining operator $G$}.
\end{itemize}
Then, the following implications hold :
$$
 (i) \Rightarrow (ii) \Rightarrow (iii) \Rightarrow (iv).$$
{  If the range $R(A\ha)$ of $A\ha$ is contained in $D(G^{-1})$, then the four conditions (i)-(iv) are equivalent}.
\enprop
\bdim 
\s{(i) $\Rightarrow$ (ii)}:
We put $K=G^{1/2}AG^{-1/2}$. We prove that $K=K \ha $.
As in Lemma \ref{lemma_onehalf2}, take $ \xi \in D(K), \, \eta \in D(K \ha )$.
Then, taking into account that $\xi \in D(G^{-1/2})$ and that, since $G^{1/2}\eta \in D(A \ha )$ and $D(A \ha )=GD(A)$,
$G^{1/2}\eta =G\zeta$ for some $\zeta \in D(A)$, we have
\begin{align*} \ip{K\xi}{\eta}&= \ip{\xi}{G^{-1/2}A \ha  G^{1/2}\eta}
= \ip{G^{-1/2}\xi}{A \ha  G^{1/2}\eta}\\
&=\ip{G^{-1/2}\xi}{A \ha  G\zeta}= \ip{G^{-1/2}\xi}{GA\zeta}\\&=\ip{G^{-1/2}\xi}{GAG^{-1/2}\eta}=\ip{\xi}{G^{1/2}AG^{-1/2}\eta}.
\end{align*}
Hence $K$ is self-adjoint.
\\[1mm]
\s{(ii)$\Rightarrow$ (iii)}: First, we prove that $A$ is symmetric in $\H(G)$; i.e., $A\subseteq A^\#$.
Indeed, if $\xi, \eta \in D(A)$, we have, by putting $\zeta=G^{1/2}\xi$ and $\varsigma = G^{1/2}\eta$,
$$ \ip{A\xi}{\eta}_G= \ip{GA\xi}{\eta}=\ip{G^{1/2} A G^{-1/2}\zeta}{\varsigma}=\ip{\zeta}{G^{1/2} A G^{-1/2}\varsigma}=\ip{\xi}{A\eta}_G.$$
Let now $\eta\in D(A^\#)\subseteq \H(G)$. Then, there exists $\eta^*\in \H(G)$, such that
$$ \ip{A\xi}{\eta}_G =\ip{\xi}{\eta^*}, \quad \forall \xi \in D(A);$$
or, equivalently,
$$ \ip{A\xi}{\widetilde{G}\eta}_G =\ip{\xi}{\widetilde{G}\eta^*}, \quad \forall \xi \in D(A).$$
Since, as noticed before, $\widetilde{G} \H(G)=D(G^{-1/2}= G^{1/2}(\H)$ and $\widetilde{G}^{1/2}\H(G)=\H$, we get the equality $\widetilde{G}=G^{1/2}\widetilde{G}^{1/2}$. Then,
$$\ip{G^{1/2}A\xi}{\widetilde{G}^{1/2}\eta}=\ip{G^{1/2}\xi}{\widetilde{G}^{1/2}\eta^*}, \quad \forall \xi \in D(A).$$
Let $\zeta:=G^{1/2}\xi$, we get
$$\ip{G^{1/2}AG^{-1/2}\zeta}{\widetilde{G}^{1/2}\eta}=\ip{\zeta}{\widetilde{G}^{1/2}\eta^*}, \quad \forall \zeta \in G^{1/2}D(A).$$
This implies that $\widetilde{G}^{1/2}\eta \in D((G^{1/2} A G^{-1/2})^*)= D(G^{1/2} A G^{-1/2})=G^{1/2}D(A).$ Thus, in particular, $\widetilde{G}^{1/2}\eta\in D(G^{-1/2}=\widetilde{G} \H(G)$. Hence $\widetilde{G}^{1/2}\eta= G^{1/2}\widetilde{G}^{1/2}\varphi$ for some $\varphi \in \H(G)$. The injectivity of $\widetilde{G}^{1/2}$, then implies that $\eta = \widetilde{G}^{1/2}\varphi\in \H$. Therefore, $G^{1/2}\eta=\widetilde{G}^{1/2}\eta\in D(AG^{-1/2})$. This, in turn, implies that $\eta \in D(A)$. In conclusion, $A$ is self-adjoint in $\H(G)$.
\\[1mm]
{ \s{(iii)$\Rightarrow$ (iv)}: Assume that $A$ is self-adjoint in $\H(G)$; i.e. $A=A^\#$. Then, by Proposition \ref{prop_214} it follows that
\begin{equation}\label{eq-domA} D(A)=\{\eta \in \H: G\eta \in D(A \ha ), \, A \ha  G\eta \in D(G^{-1}) \}.\end{equation}
Now,
$ \zeta \in GD(A)$ if and only if $G^{-1}\zeta \in D(A)$. By \eqref{eq-domA}, this is equivalent to say that $\zeta \in D(A\ha)$ and $A\ha \zeta \in D(G^{-1})$. The latter two conditions define the domain of $D(G^{-1}A\ha)$. Hence, $GD(A)=D(G^{-1}A\ha)$.
Furthermore, if $\xi \in D(A)$, Proposition \ref{prop_214} implies also that $A\xi =G^{-1}A\ha G \xi$. Then, since $A\ha G \xi\in D(G^{-1})$, by applying $G$ to both sides we conclude that $GA\xi=A\ha G\xi$.

 Finally, if $ R(A\ha)\subset D(G^{-1})$, then $D(G^{-1}A\ha)=D(A\ha)$ and (iv)$\Rightarrow$(i) is obvious.
}
\qed

\medskip

{ \berem Condition (ii) of Proposition \ref{prop_29} is equivalent to the self-adjointness of the operator $GA$ ($G$ is there bounded, with bounded inverse). So one could expect that the self-adjointness of $GA$ plays a role also when studying, as in Proposition \ref{prop_292}, the quasi-self-adjointness of $A$ in a more general context. However, it seems not to be so. One can easily prove that the condition (i) in Proposition \ref{prop_292} implies the self-adjointness of $GA$. But the self-adjointness of $GA$ seems not to be sufficient for the quasi self-adjointness of $A$.

\enrem
}
Let us now assume that {condition (ii)} of the previous proposition holds for a certain bounded metric operator $G$ and define $H:= G^{1/2}AG^{-1/2}$ on $D(H)=G^{1/2}D(A)$. Then, $H$ is self-adjoint and $HG^{1/2}\xi =G^{1/2} A\xi$ for every $\xi \in D(A)$. Clearly $G^{1/2}$ intertwines $A$ and $H$ and $A\dashv H$. We notice, on the other hand, that $G^{-1/2}$  intertwines $H$ and $A$ in the sense of Definition \ref{def_36}.

Let now $\{E(\lambda)\}$ denote the spectral family of $H$. Let $\xi \in \H$ and consider the conjugate linear functional $\Omega_{\lambda, \xi}$ defined on $D(G^{-1/2})$ by
$$
\Omega_{\lambda, \xi} (\eta)= \ip{E(\lambda)G^{1/2}\xi}{G^{-1/2}\eta}, \quad \eta \in D(G^{-1/2}).
$$
We consider here again  $D(G^{-1/2})$ as a Hilbert space, denoted by $\H(G^{-1})$, with norm $\|\cdot\|_{G^{-1}}=\|G^{-1/2}\cdot\|$ (see Section \ref{sect_intro}).
Then,
$$
|\Omega_{\lambda, \xi} (\eta)|= |\ip{E(\lambda)G^{1/2}\xi}{G^{-1/2}\eta}|\leq \|G^{1/2}\xi\|\|G^{-1/2}\eta\| =\|G^{1/2}\xi\|\|\eta\|_{G^{-1}}.
$$
Hence $\Omega_{\lambda, \xi}$ can be represented as follows
$$
\Omega_{\lambda, \xi} (\eta)= \ip{E(\lambda)G^{1/2}\xi}{G^{-1/2}\eta}=\ip{\Phi}{\eta}, \quad \eta \in D(G^{-1/2}),
$$
for a unique $\Phi\in \H(G^{-1})^\times$, the conjugate dual of $\H(G^{-1})$. It is a standard fact that $\H(G^{-1})^\times$ can be identified with $\H(G)$. We define $X(\lambda)\xi =\Phi$. Then $X(\lambda)$ is linear and maps $\H$ into $\H (G)$ continuously. One can easily prove that
$$
X(\lambda)\xi = \widetilde{G^{-1/2}}E(\lambda)G^{1/2}\xi, \quad \xi \in \H,
$$
and it obviously satisfies
$$
 \ip{X(\lambda)\xi}{\eta}=\ip{E(\lambda)G^{1/2}\xi}{G^{-1/2}\eta}, \quad \forall \,\xi \in \H, \eta \in D(G^{-1/2}).
$$

\beprop The family $\{X(\lambda)\}$ enjoys the following properties.
\begin{itemize}
\vspace*{-2mm}\item[(i)] $\dis\lim_{\lambda \to -\infty}\ip{X(\lambda)\xi}{\eta}=0; \;\lim_{\lambda \to \infty}\ip{X(\lambda)\xi}{\eta}=\ip{\xi}{\eta}, \quad \forall \,\xi \in \H, \eta \in D(G^{-1/2}).$
\vspace*{-2mm}\item[(ii)] $\dis\lim_{\lambda \downarrow \mu}\ip{X(\lambda)\xi}{\eta}= \ip{X(\mu)\xi}{\eta} , \quad \forall \,\xi \in \H, \eta \in D(G^{-1/2}).$
\vspace*{-2mm}\item[(iii)] The function $f_{\xi, \eta}:\lambda \mapsto \ip{X(\lambda)\xi}{\eta}$ is of bounded variation, for every
$\xi \in \H, \eta \in D(G^{-1/2})$, and its total variation $V(f_{\xi, \eta})$ does not exceed $\|G^{1/2}\xi\|\|G^{-1/2}\eta\|$.
\vspace*{-2mm}\item[(iv)] The following equality holds:
$$
 \ip{A\xi}{\eta}=\int_{\mb R}\lambda d\ip{X(\lambda)\xi}{\eta}, \quad \forall \,\xi \in D(A), \eta \in D(G^{-1/2}).$$
\end{itemize}
\enprop

\bdim The proof of these statements reduces to simple applications of the spectral theorem for a self-adjoint operator, similar to those given in  \cite{burnap} in an analogous situation. We simply check (iv). One has, in fact, for $\xi \in D(A)$ and $\eta \in D(G^{-1/2})$
\begin{align*}
\int_{\mb R}\lambda d\ip{X(\lambda)\xi}{\eta} &= \int_{\mb R}\lambda d\ip{E(\lambda)G^{1/2}\xi}{G^{-1/2}\eta} \\
&=\ip{HG^{1/2}\xi}{G^{-1/2}\eta}\\
& =\ip{(G^{1/2}AG^{-1/2})G^{1/2}\xi}{G^{-1/2}\eta}\\
&= \ip{A\xi}{\eta}\\[-14mm]
&\end{align*}
\qed

\smallskip

Hence, $A$ is an \emph{operator of scalar type} in a generalized sense. We notice that the representation in $(d)$ does not imply that
$\sigma(A)=\sigma(H)$.

\section{UNBOUNDED METRIC OPERATORS AND APPLICATIONS }
\label{sec:unbddmetropappl}

As explained in the introduction, one of the motivations for this paper was to study unbounded metric operators. So far, we have considered only
bounded metric operators with an unbounded inverse (i.e., quasi-invertible metric operators \cite{hoover}). Now we turn to the general case.
As expected,  results are scarce, so we mostly restrict ourselves to a discussion of various aspects.

\subsection{Unbounded metric operators}
\label{subsec:unbddmetrop}

Assume now that   $G$ is also unbounded. Let $A$ be closed and densely defined.
Then we still say that $A$ is \emph{quasi-Hermitian} if it verifies Definition \ref{quasihermitian}.
We say that $A$ is \emph{strictly quasi-Hermitian} if, in addition, $AD(A) \subset D(G)$ or, equivalently, $D(GA) = D(A)$.

In that case, indeed, $\eta\in D(A), A\eta\in D(G)$ implies $G\eta \in D(A \ha )$, so that we may write
$$
  \ip{\xi}{A \ha G\eta} = \ip{A\xi}{G\eta} = \ip{G\xi}{A\eta} = \ip{\xi}{GA\eta}, \quad \forall\, \xi, \eta \in D(A).
$$
Therefore
\be\label{eq:strqH}
A \ha G\eta = GA\eta, \quad \forall\, \eta  \in D(A).
\en
Clearly \eqref{eq:strqH} means that $A$ is quasi-Hermitian in the sense of Dieudonn\'e
 \cite{dieudonne} that is, it satisfies the relation
 $A \ha G = GA$ on the dense domain $D(A)$.

 Now, as a consequence of \eqref{eq_quasihermitian},
 the condition $D(GA) = D(A)$ is in fact equivalent to $G:D(A) \to D(A \ha )$. Thus, comparing the discussion above with the definition \ref{def:quasi-sim} of (generalized) quasi-similarity given above, we see that $A$ is  strictly quasi-Hermitian if, and only if, $A$    is  quasi-similar to $A \ha $,  i.e.,  $A\dashv A \ha $.

Of course, if $G$ is bounded, the two notions coincide.

\medskip
Although these results have some interest, they do not solve the main problem, namely, given the quasi-Hermitian operator $A$, how does one construct an appropriate metric operator $G$ { that makes it (quasi-)similar to some other, better behaved operator?} We suspect there is no general answer to the question: it has to be analyzed for each specific operator $A$.

 A partial answer may be given if one uses the formalism of \emph{partial inner product spaces} (\pip s), as described in   \cite{pip-metric,pip-book}.
Let $A =A^\times\in \mathrm{Op}(V_{J}) $ be a \emph{symmetric} operator on  an arbitrary \pip, let us say a lattice of  \hs s (LHS) $V_{J}$. Then it has been shown in Ref. \citen{pip-metric}, Prop.7.1, that, if $A$ maps $\H(G)$ into itself continuously, for some metric operator $G$, then $A$  is unitarily equivalent to a bounded operator and has a bounded  self-adjoint restriction to $\H$.

 {Now there is a sort of converse to the previous statement. Given a closed  unbounded  operator $B$ in $\H$, one may consider the self-adjoint operator
$G:= 1+(B\ha  B)^{1/2}$ and the scale $V_{\G}$  built on the powers of $G^{1/2}$ (this is essentially the only \pip\ one can build intrinsically from $B$ alone), namely,   $V_{\G}:= \{\H_{n}, n \in \ZN \}$,
where $\H_{n} =  D(G^{n/2}),  n\in \NN$, with a norm equivalent to the graph norm, and $ \H_{-n} =\H_{n}^\times$:
\be\label{eq:scale}
 \ldots\subset\; \H_{2}\; \subset\;\H_{1}\; \subset\; \H \; \subset\; \H_{-1} \subset\; \H_{-2} \subset\; \ldots
\en
Thus $\H_{1} =   \H(G)$ and $\H_{-1} =  \H(G^{-1})$.
Then $T=G^{-1/2}$ is a bounded metric operator. Hence,  according to
Proposition \ref{prop410}, $B$ is quasi-Hermitian with respect to $T$ if and only if  $A_0= TB$ is symmetric in $\H$. Next, since $D(A_0)$ is dense in $\H$,
$A_0 $ defines a unique symmetric operator $A=A^\times$ in the scale $V_{\G}$.}

 Another open question is the following. Given two closed operators $A,B$, under which conditions are they (quasi-)similar to each other? According to the discussion so far, these conditions will be of a spectral nature, such as equality of the spectra or of some part of the spectra.

\subsection{Pseudo-Hermitian Hamiltonians}
\label{subsect_psH}

It is well-known that metric operators appear routinely in the so-called pseudo-Hermitian quantum mechanics \cite{bender},  but in general only bounded ones are considered. However, unbounded metric operators have been discussed in some recent work \cite{bag-zno2,kretsch,mosta2}. The question is, how do such operators fit in the present formalism?

 The starting point of   \cite{mosta2} is a so-called reference \hs\ $\H$ and a quasi-Hermitian operator\footnote{
The author of  \cite{mosta2} calls this a $G$-pseudo-Hermitian operator, but in fact it is simply a quasi-Hermitian operator, in the original sense of 
Dieudonn\'e \cite{dieudonne},  but unbounded.}
$H$ on $\H$, which means  there exists a possibly  unbounded metric operator $G$  satisfying the relation
\be\label{eq:psH}
H \ha  G = G H.
\en
This operator $H$ is taken as the non-self-adjoint (but $\P\T$-symmetric) Hamiltonian of a quantum system.

In the relation \eqref{eq:psH}, the two operators are assumed to have the same  domain, $D(H \ha  G)=D(GH)$,
which is supposed to be dense. This condition is not necessary, however, if we assume that $H$ is quasi-Hermitian in the sense of Definition \ref{quasihermitian}.
This means indeed that $D(H) \subset D(G)$ and
\be\label{eq_quasihermitian4}
\ip{H\xi}{G\eta}= \ip{G\xi}{H\eta},  \;\; \forall\, \xi, \eta \in \D(H).
\en
If $G$ is bounded, we get $H\dashv H\ha$ and {then $H$ will be quasi-self-adjoint if one of the conditions of Proposition \ref{prop_292} is satisfied; for instance if $G^{1/2} H G^{-1/2}$ is self-adjoint}.

On the other hand, if $G$ is unbounded and if we assume that $H$ is strictly quasi-Hermitian, in the sense of Section \ref{subsec:unbddmetrop}, we still have $H\dashv H\ha$, but we cannot conclude.
However, if in addition
$G^{-1}$ is bounded, we get $G^{-1}H \ha  G\eta = H\eta, \; \forall \,\eta\in \D(H)$, which is a more restrictive form of similarity.

Finally, if one assumes, in addition,  that the quasi-Hermitian operator $H$  possesses a (large) set $\D$ of vectors 
which are  analytic in the norm  $\norm{G}{\cdot}$ and are contained in $D(G)$ \cite{barut-racz,nelson}, 
the construction given in  \cite[Sec.6]{pip-metric}, can be performed. In a nutshell, one endows $\D$ with the norm $\norm{G}{\cdot}$ and takes the completion $\H_G$, which is a closed subspace of $\H(G)$. Next, $G^{1/2}$ extends to an isometry $W =  \H_G \to \H$, with closed range $\H_{\rm phys}$.
Then $H$ extends to a self-adjoint operator $\ov H$  in $\H_G$ and the operator $h=  W\, \ov H\,W^{-1}$ is self-adjoint in $\H_{\rm phys}$\, .

If $\D$ is dense in $\H$, then $\H_G= \H(G)$, $\H_{\rm phys}= \H$  and $W = G^{1/2}$ is unitary from $\H(G)$ onto $ \H$. Thus $H$ is unitarily equivalent to the self-adjoint operator $h$.

\subsection{An example}
\label{subsect_ex}

A beautiful example of the situation just analyzed has been given recently by Samsonov \cite{samsonov}, namely, the second derivative on the positive half-line
(this example stems from Schwartz \cite{schwartz}):
\be\label{eq-H}
H = -\frac{\ud ^2}{\ud  x^2}, \quad x \geq 0,
\en
with domain
\be\label{eq-DH}
D(H) = \{\xi \in L^2(0,\infty) : \xi ''  \in L^2(0,\infty), \xi'(0) + (d+i b)\xi(0) = 0\}.
\en
For $d<0$, this operator has a purely continuous spectrum. Its adjoint $H\ha$ is given again by \eqref{eq-H}, on the domain
$D(H\ha)$ defined as in \eqref{eq-DH}, with $b$ replaced by $-b$.

Next introduce the unbounded operator
\be
G = -\frac{\ud ^2}{\ud  x^2} -2 ib \frac{\ud }{\ud  x} + d^2 +b^2,
\en
on the domain $D(G) = D(H)$. Then a direct calculation shows that $G$ is self-adjoint, strictly positive and invertible, i.e., it is a metric operator. Since its spectrum is 
$\sigma(G) = \sigma_c(G) = [d^2, \infty)$, it follows that  $G^{-1}$ is bounded.

Since both $H$ and $G$ are second order differential operators, an element of the domain $D(GH)$ should have a square integrable fourth derivative. Hence one defines
\be\label{eq:tildeD}
\widetilde{D}(H) = \{\xi \in D(H) : \xi^{\rm (iv)} \in L^2(0,\infty)\} \subset D(H).
\en
and $ \widetilde{H} = H \up \widetilde{D}(H)$.
Then the analysis of \cite{samsonov} yields the following results:
\bee
\vspace*{-2mm}\item[(i)] $H$ is quasi-Hermitian in the sense of Definition \ref{quasihermitian}, that is, it satisfies the relation \eqref{eq_quasihermitian4}
on $ D(G)=  D(H)$.

\vspace*{-2mm}\item[(ii)] $G$ maps $\widetilde{D}(H) $ into $D(H\ha)$.

\vspace*{-2mm}\item[(iii)] $H$ is quasi-Hermitian in the sense of Dieudonn\'e, that is, $GH = H\ha G$ on  the  dense domain $\widetilde{D}(H)$.
We have to restrict ourselves to $\widetilde{D}(H)$ because of the requirement on the fourth derivative.

\vspace*{-2mm}\item[(iv)] The operator $h = G^{1/2} H G^{-1/2} = G^{-1/2} H\ha G^{1/2}$ is self-adjoint on the domain 
$D(h) = \{\eta=  G^{1/2}\xi, \, \xi \in\eta \}$.
\ene
In conclusion, by (i) and (ii), we   get $\widetilde{H}\dashv H\ha$. 

In fact, one could use as metric operator $T:=G^{-1}$, which is bounded, with unbounded inverse $T^{-1}=G$, a more standard situation, and get similar results.
However, since we don't know the operator $T$ explicitly, this is of little use

\section*{ACKNOWLEDGMENTS }

Part of this work was done in the Department of Applied Mathematics, Fukuoka University. We both thank Prof. A. Inoue for his kind hospitality.

\end{document}